\documentclass[debug,overfull]{epl}

\usepackage{amssymb}

\title{Multifractal properties of growing networks}
\shorttitle{Multifractal properties of growing networks}
\author{S.N. Dorogovtsev\inst{1,2}\thanks{E-mail: \email{sdorogov@fc.up.pt}} 
\and J.F.F. Mendes\inst{1}\thanks{E-mail: \email{jfmendes@fc.up.pt}} 
\and A.N. Samukhin\inst{2}\thanks{E-mail: \email{alnis@samaln.ioffe.rssi.ru}}}
\institute{
  \inst{1} Departamento de F\'{\i}sica and Centro de F\'{\i}sica do Porto, Faculdade
de Ci\^{e}ncias, Universidade do Porto,
Rua do Campo Alegre 687, 4169-007 Porto, Portugal\\
  \inst{2} A.F. Ioffe Physico-Technical Institute, 194021 St. Petersburg, Russia
}

\begin{document}

\maketitle

\begin{abstract}
We introduce a new family of models for growing networks. In these
networks new edges are attached preferentially to vertices with higher
number of connections, and new vertices are created by already existing
ones, inheriting part of their parent's connections. We show that
combination of these two features produces multifractal degree
distributions, where degree is the number of connections of a vertex. An
exact multifractal distribution is found for a nontrivial model of this
class. The distribution tends to a power-law one, $\Pi \left( q\right) \sim
q^{-\gamma }$, $\gamma =\sqrt{2}$ in the infinite network limit.
Nevertheless, for finite networks's sizes, because of multifractality,
attempts to interpret the distribution as a scale-free would result in an
ambiguous value of the exponent $\gamma $.
\end{abstract}

%\section{Section title}
Networks of various kinds, such as the World Wide Web, citation network of
scientific papers, social networks, neural networks, etc. (see \cite
{ws98,rs98,hppl98,ajb99,ba99,nmej00,cmp00,bb00}) are very popular
objects of studies nowadays. There crucial role is played by the \emph{%
degree distribution function }(DDF) $\Pi \left( q\right) $, where the degree 
$q$ is the number of connections of a vertex (sometimes it is called
connectivity). Data, obtained from the observations of many existing
networks were interpreted as if they are \emph{scale-free}, that is degrees
of their vertices were distributed according to $\Pi \left( q\right) \sim
q^{-\gamma }$ \cite{ba99}. Such networks have some specific
properties, as compared, e.g., to classic random graphs \cite{er60}, where
degree distribution follows a Poisson's law. In particular, they are
extremely resilient to random breakdowns \cite{ajb00,cnsw00,ceah00}. 

To describe a scale-free growing network, the mechanism of \emph{%
preferential linking} \cite{ba99} was proposed. This principle is similar to
the one introduced in the well known Simon model \cite{sha55}, used to
explain power-law distributions in various social and economic systems. In
fact, all these models belong to the class of stochastic multiplicative
processes \cite{stoch}. Here we show that a natural generalization of
studied models of evolving networks leads to multifractal degree
distribution with much reacher properties than a power-law one.

In general, as network is growing, two parallel processes take place: (i)
New edges are formed between vertices. They attach preferentially to
vertices of a growing network with a high number of connections. (ii) New
vertices appear. Several types of such a models were proposed \cite
{ba99,dms00,krl00,dm00b,bb00}, that differ in details of how new vertices
and edges are entered into the network. They produced scale-free networks
with the $\gamma $ exponent of the DDF, either in the range $\left( 2,\infty
\right) $ \cite{ba99,dms00,krl00}, which means, that the average degree of a
vertex remains finite, or $\gamma <2$ \cite{dm00b,fs01}, when number of
edges is growing faster then of vertices.

The questions are: whether all the scenarios of the evolution of networks,
produced by preferential linking, lead to scale-free DDF? What other types
of DDF do arise? These questions were considered in several recently
published papers \cite{dms00,krl00,dm00b,ab00,bb00}. However, in 
%%all the previous  
most of the 
models, where the idea of preferential linking was used, it was
assumed, that new vertices appear with the same properties, independent of
the state of network at this moment, i.e., all vertices are born equal. 
(Nevertheless, see the counter-example of a ``breathing network'' in Ref. \cite{sk99}). 
One
could say, that new vertices are created by some external source, whose
properties are independent of the network's current state. Here we put
forward a different concept: \emph{new vertices are born with random
properties, which reflects the state of the network at the moment of birth}.
In this respect, one could say, that they are created by the network itself.

\begin{figure}[tbp]
\onefigure[scale=0.45]{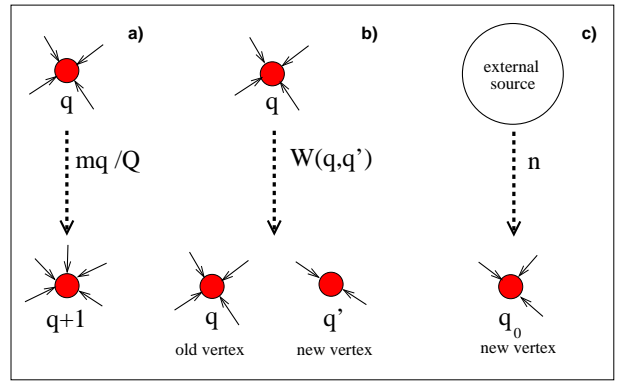}
\caption{Elementary processes, defining the evolution of a network. (a) ---
creation of new link, pointing at a node. Probability of this process per
unit time is $mq(t)/Q(t)$. (b) --- creation of new node by a randomly
chosen member of the network. Its probability is $W\left(q,q^{\prime}%
\right)=n{q\choose q^{\prime} }c^{q^{\prime}}(1-c)^{q-q^{\prime}}$. 
(c) --- creation of new
node by external sourse. This process has the probability $n$.}
\label{fig1}
\end{figure}

Our model may be formulated as follows: (i) New edges are introduced between
existing vertices (see Fig.\ref{fig1}a). This happens with probability $m$
per unit time. We assume the linear preferential linking: the probability, a
new edge to point to a vertex $i$ of in-degree $q_i$ is $p_i=q_i/Q$. Here $%
Q\left( t\right) =\sum_iq_i$ is total degree (number of edges) of the
network. We shall look only at the in-degrees (numbers of incoming edges) of
vertices, putting aside the question of what vertices these edges come from.
For example, one may assume that adge edge comes from any vertex with equal
probability. (ii) New vertices appear with probability $n$ per unit time
(Fig.\ref{fig1}b) (in the following we assume $n=1$, which can always be
done by the proper choice of the time scale). We suppose that every new
vertex has a parent --- some vertex, randomly chosen among $N\left( t\right) 
$ existing ones. Degree of a daughter vertex is assumed to be a random
number, whose distribution depends upon parent's degree. Namely, we assume,
that every edge, pointing to the parent with some probability $c$ is
inherited by the daughter. We emphasize, that here inheritance means copying
of edges, --- parent loses nothing, simply some new edges are created,
pointing at its heir. If parent's degree is $q$, the probability for its
heir to have the degree $q_1\le q$ is ${q\choose q_1} c^{q_1}\left( 1-c\right)
^{q-q_1}$.

It is the way of introducing new vertices into the network, that makes our
model differ significantly from previously studied ones. Usually it is
assumed, that new vertices have some preset properties, which are
independent of the current state of the network. For example, in most cases
new vertices are entered with some given initial degree $q_0$ (see Fig.\ref
{fig1}c). In our model new vertices are being born by network's components
themselves, and with properties (initial degrees), reflecting the current
state of the network.

DDF may be introduced as: 
$\Pi \left( t,q\right) =\left\langle [1/N(t)]
\sum_{j=1}^{N\left( t\right) }\delta _K\left[ q_j\left( t\right) -q\right]
\right\rangle $, 
where $\delta _K\left( q-q^{\prime }\right) $ is the notation for
Kronecker's $\delta $-symbol. Averaging is over the statistical ensemble of
networks, whose evolution is governed by the above described rules. Here $%
N\left( t\right) $ is an integer, growing in time in a random fashion, ---
introduction of new vertices (as well as new links) may be considered as a
random process (e.g., Poissonian one). Its average is $\bar{N}=\int
dt\,n\left( t\right) =t$. For $t\gg 1$, this function may be shown to obey
the following master equation: 
\begin{equation}
\frac \partial {\partial t}t\Pi \left( t,q\right) +\frac m{\bar{q}\left(
t\right) }\left[ q\Pi \left( t,q\right) -\left( q-1\right) \Pi \left(
t,q-1\right) \right] =\sum_{q^{\prime }=q}^\infty {q^{\prime }\choose q}
c^q\left( 1-c\right) ^{q^{\prime }-q}\Pi \left( t,q^{\prime }\right) \,.
\label{20}
\end{equation}
Here $\bar{q}\left( t\right) =\bar{Q}\left( t\right) /t$ is the average
degree of network's vertices. The above equation may easily be derived, but,
as one can see, it is a direct generalization of the master equation,
introduced in \cite{dms00}. The only difference is that here in the rhs,
instead of $\delta _K\left( q-q_0\right) $, which corresponded to vertices,
born by some external source with definite degree, (Fig.\ref{fig1}c), we
have a term, reflecting DDF of a network, reached at the given moment. All
terms in Eq.(\ref{20}) have transparent physical meaning, reflecting balance
between income and outcome within the number of vertices with degree $q$, $%
t\Pi \left( t,q\right) $ (note, that $t$ variable is rather the total number
of vertices in the network, then the time). From Eq.(\ref{20}) one can
easily obtain: $\bar{q}\left( t\right) =m/[(1-c)]\left( 1+bt^{c-1}\right) $,
where $b$ is a constant of integration. Then, in the limit of large size $t$%
, $m/\bar{q}\left( t\right) $ may be replaced with $1-c$.

Coefficient $c$ reflects the ``succession right'' of a given network. In
principle, it can also be a random number, characterized by some
distribution density $h\left( c\right) $. The general evolution equation
acquires much simpler form, if to pass to a new representation of the DDF,
which is a bit modified Z-transform: 
$
\Pi \left( t,q\right) \rightarrow \Phi \left( t,y\right) =\sum_{q=0}^\infty
\Pi \left( t,q\right) \left( 1-y\right) ^q=\left\langle [1/N(t)]
\sum_{j=1}^{N\left( t\right) }\left( 1-y\right)
^{q_j}\right\rangle
$.
Then the evolution equation becomes: 
\begin{equation}
\frac \partial {\partial t}t\Phi \left( t,y\right) -\left( 1-\bar{c}\right)
y\left( 1-y\right) \frac{\partial \Phi \left( t,y\right) }{\partial y}%
-\int_0^1dc\,h\left( c\right) \Phi \left( t,cy\right) =0\,.  \label{40}
\end{equation}
It is to be supplied with the initial condition: $\Phi \left( t_0,y\right)
=\Phi _0\left( y\right) $, $t_0\gg 1$ (recall, that the equation is valid
for $t\gg 1$). After rescaling of the size variable $t\rightarrow t/t_0$ the
initial condition becomes $\Phi \left( 1,y\right) =\Phi _0\left( y\right) $.

General solution of Eq.(\ref{20}) may be found in the continuous
approximation, when it is assumed, that $\Pi \left( t,q\right) $ is a slowly
varying function of $q$. Then this equation, taking into account the above
introduced randomness of $c$, takes the form: 
\begin{equation}
\frac \partial {\partial t}t\Pi \left( t,q\right) +\left( 1-\bar{c}\right)
\frac \partial {\partial q}q\Pi \left( t,q\right) -\int_0^1\frac{dc}%
c\,h\left( c\right) \Pi \left( t,\frac qc\right) =0\,.  \label{50}
\end{equation}
It can easily be solved after transition to Mellin's representation with
respect to $q$: 
$
\Xi \left( t,\xi \right) =\int_0^\infty dq\,\Pi \left( t,q\right) q^{\xi -1}
\label{60}
$.
Note, that $\Xi \left( t,n+1\right) =M_n\left( t\right) $ are the moments of
the distribution. The solution is: 
\begin{equation}
\Xi \left( t,\xi \right) =\Xi _0\left( \xi \right) t^{\tau \left( \xi
-1\right) }\,,\;\tau \left( \xi \right) =\left( 1-\bar{c}\right) \xi -1+\chi
\left( \xi +1\right) \,,\;\chi \left( \xi \right) =\int_0^qdc\,h\left(
c\right) c^{\xi -1}\,.  \label{70}
\end{equation}

One can see, that moments of the distribution scale with network as $M_n\sim
t^{\tau \left( n\right) }$, where $\tau \left( n\right) $ is a nonlinear
function of $n$. These distributions are usually referred to as \emph{%
multifractals }\cite{mbb74,hp83,hjkps86}, as distinct from fractal ones,
where $\tau \left( n\right) $ depends on $n$ linearly, $\tau \left( n\right)
=\left( n-1\right) D$, $D$ is called fractal's dimensionality. Such a
distributions were found in many objects, ranging from localized excitations
in disordered solids to the distribution of matter in the Universe.
Mutifractals may be thought of as a statistical mixture of fractals with
different dimensionalities. Indeed, if to introduce the $f\left( \alpha
\right) $\emph{-spectrum} of dimensionalities $\alpha $, then a mixture of $%
\alpha $-dimensional fractals with statistical weights $g\left( \alpha
\right) t^{-f\left( \alpha \right) }$, ascribed to $\alpha $-dimensional
fractal, yields the distribution moments $M_n=\int d\alpha \,g\left( \alpha
\right) t^{\alpha \left( n-1\right) -f\left( \alpha \right) }$. At large $t$
they scale with network's size as $M_n\sim t^{\tau \left( n\right) }$. The
prefactor $g\left( \alpha \right) $ gives only the coefficients of
proportionality, and is not relevant. Functions $f\left( \alpha \right) $
and $\tau \left( n\right) $ are connected by the Legendre transform: $\tau
\left( n\right) +f\left( \alpha \right) =\alpha n$, $n=df/d\alpha $, $\alpha
=d\tau /dn$.

As an example, let us consider the case of the homogeneous distribution of
the inheritance coefficient $c$, $h\left( c\right) =\theta \left( c\right)
\theta \left( 1-c\right) $. The motivation for such a choice is that, as we
shall see below, in this case an exact solution is possible without
continuous approximation. We have $\chi \left( \xi \right) =1/\xi $, and $%
\tau \left( \xi \right) =\xi /2-1+1/\left( \xi +1\right) $. This gives
multifractal $f\left( \alpha \right) $-spectrum to be: $f\left( \alpha
\right) =(1-\sqrt{1/2-\alpha })^2$, where $-\infty <\alpha <1/2$. Note, that
it includes negative dimensionalities (see \cite{mbb89}).

If we set the initial condition for Eq.(\ref{50}) as $\Pi \left( 1,q\right)
=\Pi _0\left( q\right) =\delta \left( q-q_0\right) $, then the solution
is a Green's function $\Pi \left( t;q,q_0\right) $, whose expression is: 
\begin{eqnarray}
&&\Pi \left( t;q,q_0\right) =\frac 1{q_0t^{3/2}}\int_{-i\infty +\delta
}^{+i\infty +\delta }\frac{d\xi }{2\pi i}\left( \frac{q_0}q\right) ^\xi
t^{\xi /2+1/\xi }  \nonumber \\
&=&\frac{\theta \left( q_0\sqrt{t}-q\right) }{q_0t^{3/2}}\sqrt{\frac{\ln t}{%
\ln \left( q_0\sqrt{t}/q\right) }}I_1\left( 2\sqrt{\ln t\,\ln \left( q_0%
\sqrt{t}/q\right) }\right) +\frac 1t\delta \left( q-q_0\sqrt{t}\right) \,,
\label{80}
\end{eqnarray}
where $I_1$ is the modified Bessel's function. If $t$ is large enough, the
above formula everywhere, except the tail region $q\approx q_0\sqrt{t}$, may
be replaced by: 
\begin{equation}
\Pi \left( t;q,q_0\right) \approx \frac{2^{-1/4}}{q_0t^{3/2}\sqrt{\pi \ln t}}%
\exp \left( 2\sqrt{\ln t\,\ln \left( q_0\sqrt{t}/q\right) }\right) \,\,.
\label{90}
\end{equation}
At $t\rightarrow \infty $ this expression is asymptotically equal to: 
\begin{equation}
\Pi \left( t;q,q_0\right) \approx \frac{2^{-1/4}}{q_0\sqrt{\pi }}\left( 
\frac{q_0}q\right) ^{\sqrt{2}}\frac{t^{-3/2+\sqrt{2}}}{\left( \ln t\right)
^{1/2}}\,\,.  \label{100}
\end{equation}
However, this latter result is valid only if $\ln \left( q/q_0\right) \ll 
\sqrt{\ln t}$. At $t\rightarrow \infty $ this region is small as compared
with the one of the validity of Eq.(\ref{90}), which is $\ln q\lesssim
\left( 1/2\right) \ln t$. This means that, in spite of that the distribution
asymptotically assumes a scale-free form, $\sim q^{-\gamma }$ with $\gamma =%
\sqrt{2}$, at finite network's sizes this is true only within rather
restricted region of degrees, small compared with the upper cut-off $q_0%
\sqrt{t}$. However, it is extremely important, that one has to compare not
the sizes of the regions themselves, but their logarithms. This ratio is of
the order of $1/\sqrt{\ln t}$. Even for the largest known network, the World
Wide Web, whose size is $\sim 10^9$, this value is of the order of $0.2$.
These features make the analysis of the distribution in terms of scale-free
functions ambiguous, the value of the exponent $\gamma $ becomes dependent
on the choice of the degree region, from which it is extracted. The
(negative) slope of a log-log plot of the distribution steadily grows with $%
\ln q$ from $\sqrt{2}$ at $\ln q\lesssim \sqrt{\ln t}$, until the
exponential cut-off is reached at $\ln q\sim \ln t$.

There is another peculiar feature of the distribution (\ref{80}), obtained
in the continuum approximation. Namely, for any $q_c>0$ at large enough $t$
the fraction of vertices with $q>q_c$ scales with $t$ as $t^{-3/2+\sqrt{2}}$%
. This means, that the distribution concentrates within the region of small $%
q$, where continuum approach is invalid. Fortunately, for the homogenous
distribution of inheritance coefficient $c$, the exact solution is possible.
Indeed, for the homogeneous $h(c)$, after application of the operator $%
\partial _y(y\cdot )$, Eq. (\ref{40}), may be reduced to a linear partial
differential equation. This equation, after the Mellin's transformation with
respect to time, $\psi (\eta ,y)=\int_1^\infty dt\,t^{\eta -1}\Phi (t,y)$,
takes the form,

\begin{equation}
y^2\!\left( 1\!-\!y\right) \partial _y^2\psi +y\left( 2\eta -3y\right)
\partial _y\psi +2\eta \psi =\!-2\partial _y\!\left( y\Phi _0\right) \!,
\label{110}
\end{equation}
where $\Phi _0\left( y\right) $ is the initial distribution in the
Z-representation. For the Green's function it is: $\Phi _0\left( y\right)
=\left( 1-y\right) ^{q_0}$. Let us denote Mellin's time-transform of the
Green's function as $\Psi \left( \eta ;q,q_0\right) $. Eq. (\ref{110}) may
be reduced to an inhomogeneous hypergeometric one after the substitution $%
\psi \left( y\right) =y^\zeta \chi \left( y\right) $, where $\zeta $ is one
of the roots of the characteristic equation: $\zeta ^2-\left( 1-2\eta
\right) \zeta +2\eta =0$. Here we present the result in terms of $\Psi
\left( \eta ;q,k\right) $, which is the $q$-th term of Taylor's series of $%
\psi \left( y\right) $ around the point $y=1$. After lengthy calculations,
we obtain

\begin{eqnarray}
\Psi \left( \eta ;q,k\right) &=&\left\{ 
\begin{array}{l}
-\phi _1\left( \zeta _1,k\right) \frac{\phi _2\left( \zeta _1,q\right) -\phi
_2\left( \zeta _2,q\right) }{\zeta _1-\zeta _2}\,,\;k>q>0; \\[7pt] 
-\frac{\phi _1\left( \zeta _1,k\right) -\phi _1\left( \zeta _2,k\right) }{%
\zeta _1-\zeta _2}\phi _2\left( \zeta _1,q\right) \,,\;q\ge \,k>0;
\end{array}
\right. \;  \nonumber \\
\Psi \left( \eta ;0,k\right) &=&-\phi _1\left( \zeta _1,k\right) \;;\;\Psi
\left( \eta ;q,0\right) =-\frac{\delta \left( q\right) }\eta \;,  \label{120}
\end{eqnarray}
where $\zeta _1$ is the root of the characteristic equation, which is
positive for $\mathrm{Re}\,\eta <0$, $\zeta _2$ is the other root, and
functions $\phi _{1,2}$ may be expressed as:

\begin{eqnarray}
\phi _1\left( \zeta ,q_0\right) &=&\frac{4q_0\Gamma (\zeta )\Gamma (2+\zeta
)/(1-\zeta )}{\Gamma \left( 1\!+\!\zeta +\frac 2{1+\zeta }\right) \Gamma
\left( 2\!+\!\zeta -\frac 2{1+\zeta }\right) }\!  \nonumber \\
&&\int_0^1\!dz\,z^{1-2/\left( 1+\zeta \right) }\left( 1-z\right)
^{q_0-1}\,_2F_1(\!1-\frac 2{1\!+\!\zeta },1-\frac 2{1\!+\!\zeta };2+\zeta
-\frac 2{1\!+\!\zeta };z),  \label{130}
\end{eqnarray}
\vspace{-14pt} 
\begin{eqnarray}
\phi _2\left( \zeta ,q\right) &=&\frac{\Gamma \left( 1+\zeta +\frac
2{1+\zeta }\right) \Gamma \left( \frac 2{1+\zeta }-\zeta \right) }{\Gamma
\left( \frac 2{1+\zeta }-1\right) \Gamma \left( \frac 2{1+\zeta }+1\right) }%
\frac{\sin \pi \zeta }\pi  \nonumber \\
&&\int_0^\infty dy\,y^\zeta \left( 1+y\right) ^{-q-1}\,_2F_1\left( \zeta
,2+\zeta ;2+\zeta -\frac 2{1+\zeta };-y\right) \,.  \label{140}
\end{eqnarray}
Here $_2F_1$ is the hypergeometric function.

Two formulas may be derived from Eqs.(\ref{120}-\ref{140}),
which indicate: i) that the results of the continuous
approximation are valid for large $q$, except for some minor corrections,
and ii) that concentration of the distribution at low degree values is to be
interpreted as in the large size limit, almost all vertices have zero
degree. In the large $q$, large $t$ limit, the following expression may be
written for the Green's function:

\begin{equation}
\Pi \left( t;q,q_0\right) \approx \omega ^{\prime }g\left( q_0\right) \frac{%
\ln \left( aq\right) }{\left( t\ln t\right) ^{3/2}}\exp \left[ 2\sqrt{\ln
t\ln \left( \sqrt{t}/q\right) }\right] \,,\;g\left( q_0\right) =\left|
\left. \frac{\partial \phi _1\left( \zeta ,q_0\right) }{\partial \zeta }%
\right| _{\zeta =\sqrt{2}-1}\right| \,,  \label{150}
\end{equation}
where: $\omega ^{\prime }=0.0823\dots $ and $a=0.840\dots $ It differs from
Eq.(\ref{90}) in some details, but the main conclusions remain the same. The
formula (\ref{100}) is reproduced, apart from the appearance of the
additional multiple $\ln q/\ln t$ and from different numerical factors.
Also, the dependence on the initial degree value $q_0$ is different (one may
reproduce completely continuous approximation result (\ref{80}) from the
exact solution, if to assume the limit $q\rightarrow \infty $, $%
q_0\rightarrow \infty $, $q/q_0$ and $t$ fixed). Anyway, apart from some
power of logarithm, DDF values for any $q>0$ decay as $t^{-3/2+\sqrt{2}}$
with network's size increase. For $q=0$, we have:

\begin{equation}
\Pi \left( t;0,q_0\right) \approx 1-g\left( q_0\right) \frac{t^{-3/2+\sqrt{2}%
}}{2^{1/4}\sqrt{\pi }\ln ^{3/2}t}\,.  \label{160}
\end{equation}

One can see from Eq. (\ref{160}) that at long times the fraction of
zero-degree vertices tends to $1$. These vertices do not have incoming edges
but only outgoing ones (recall, that degree, in the present paper, is
defined as in-degree, the number of incoming edges). They are passive
constituents of the network, i.e., their degree remains unchanged all the
time. Although the fraction of active vertices (with non-zero degree) tends
to zero as the network grows, their total number \emph{increases} with time
as $t^{\sqrt{2}-1/2}/\ln ^{3/2}t$. One can introduce the DDF of active
vertices, $\Pi _1\left( t;q,q_0\right) $. It follows from Eqs. (\ref{150})
and (\ref{160}), that in the large size limit, it tends to the distribution, 
$\Pi _1\left( q,t\right) $, which does not depend on the initial conditions:

\begin{equation}
\Pi _1\left( t,q\right) \equiv \left( 1-\delta _{q0}\right) \frac{\Pi \left(
t;q,q_0\right) }{1-\Pi \left( t;0,q_0\right) }\rightarrow \omega t^{-\sqrt{2}%
}\ln \left( aq\right) \exp \left[ 2\sqrt{\ln t\ln \left( \sqrt{t}/q\right) }%
\right] \,.  \label{170}
\end{equation}
$\omega =0.174\dots $. One can obtain from Eq. (\ref{170}) the following
asymptotic expression at $t\rightarrow \infty $:

\begin{equation}
\Pi _1\left( q\right) =\frac \omega {q^{\sqrt{2}}}\ln \left( aq\right) \,.
\label{180}
\end{equation}
Apart of the logarithmic factor, this is a power-law (scale-free) DDF with
exponent $\gamma <2$. However, the range of validity is small, $\ln q\ll 
\sqrt{\ln t}$, and the moments of the distribution are defined by its
behavior outside ``scale-free'' region. More general expression (\ref{170})
is valid if $q<q_0\sqrt{t}$, at $q\approx q_0\sqrt{t}$ there exist an
exponential cut-off, which replaces the $\delta $-functional term at the
point of abrupt cut-off, found in continuous approximation (Eq.(\ref{80})).

In conclusion: natural generalization of network's evolutionary dynamics
with preferential linking was suggested. We have found that evolving
networks with preferential linking, in which new vertices are born by
previously existing ones, and inherit edges from their parent, arrive at the
multifractal degree distribution. For a specific model, exact large-size
solution was found. Degree distribution tends to nearly power-law type in
the infinite network limit, $\Pi (q)\sim q^{-\gamma }\ln q$, $1<\gamma <2$
(specifically, $\gamma =\sqrt{2}$ for the exactly solvable model).
Nevertheless, for finite network the region $q<\exp \left( \sqrt{\ln t}%
\right) $ where this behavior may be observed, is rather small as compared
to one of relevance, $q<q_0\sqrt{t}$. With probability close to $1$ a
vertex, randomly chosen among ones with nonzero degree (active vertex), has
its degree within the power-law dependence region. However, distribution's
moments will be defined by the broad large-degrees tail, containing a small
fraction of vertices. Results, obtained here for the model with a particular
(homogeneous) distribution of inheritance coefficient, may easily be
generalized to a more generic case of any distribution.

In most of empirical scale-free distributions of degrees the exponent $%
\gamma $ was found to be $\gamma >2$. Also, in most models, $\gamma >2$,
that means that the average degree remains finite as network grows. Note
that the degree distributions with $\gamma <2$ were obtained analytically
and by simulation for networks with accelerating growth \cite{dm00b}, where $%
\bar{q}\rightarrow \infty $ as $t\rightarrow \infty $. In the present case,
the number of edges per \emph{active} site also grows with time in a power
law manner, i.e., as $t^{3/2-\sqrt{2}}\ln ^{3/2}t$. However, the case of
multifractal distribution is essentially different from a scale-free one.
Here the distribution can not be described, using one, or finite set of
scaling exponents. In particular, the attempt to analyze the distribution in
terms of a scale-free one would lead to the value of the exponent $\gamma $,
dependent on the method of analysis.

\acknowledgments
SND thanks PRAXIS XXI (Portugal) for a research grant PRAXIS
XXI/BCC/16418/98. JFFM and SND were partially supported by the project
POCTI/1999/Fis/33141. ANS acknowledges the NATO program OUTREACH for
support. We also thank V.V. Bryksin, A.V. Goltsev and B.N. Shalaev for many
useful discussions.

\end{document}